\documentclass[a4paper, oneside, twocolumn, notitlepage, 10pt]{extarticle_ecoc}
\usepackage{ecoc}

\begin{document}
\selectlanguage{english}    


\title{Capitalizing on Next-Generation Optical Communication Systems with Proactive Multi-Period Network Planning}%


\author{
    Jasper Müller\textsuperscript{(1, 2)}, 
    Sai Kireet Patri\textsuperscript{(1, 2)}, 
    Gabriele Di Rosa\textsuperscript{(1)}, 
    Achim Autenrieth\textsuperscript{(1)}, \\
    Jörg-Peter Elbers\textsuperscript{(1)} 
    and Carmen Mas-Machuca\textsuperscript{(3)}
}

\maketitle                  


\begin{strip}
 \begin{author_descr}

   \textsuperscript{(1)} Adtran, Fraunhoferstr. 9a, 82152 Martinsried/Munich, Germany,
   \textcolor{blue}{\uline{jmueller@adva.com}}

   \textsuperscript{(2)} Chair of Communication Networks, Technical University of Munich, Arcisstr. 21, Munich, Germany
   
   \textsuperscript{(3)} Chair of Communication Networks, Universit\"at der Bundeswehr M\"unchen, Germany

 \end{author_descr}
\end{strip}

\setstretch{1.1}
\renewcommand\footnotemark{}
\renewcommand\footnoterule{}

\newcommand{\TXOSNR}{OSNR\textsubscript{TX}}


\begin{strip}
  \begin{ecoc_abstract}
    Optical transport network operators typically follow a pay-as-you-grow strategy for their network deployment. We propose a proactive multi-period planning approach based on heuristic network planning, supporting this deployment strategy while enabling efficient network utilization through next-generation technology. We report 60\% less provisioned lightpaths. \textcopyright2023 The Author(s)
  \end{ecoc_abstract}
\end{strip}


\section{Introduction}
As fifth-generation~(5G) mobile communication systems are being rolled out extensively, standardization and definition efforts for sixth-generation~(6G) communication are already underway. Consequently, exponential growth in data traffic demands is challenging optical transport network (OTN) operators in the coming years~\cite{Jiang:19}.

Recent developments in optical communication hardware allow for improved resource efficiency in optical networks. These include high-baud rate, modulation rate-adaptive bandwidth-variable transceivers (BVTs) \cite{Jannu}, multi-wavelength transponders~\cite{COMBS} and multi-band systems~\cite{Emmerich:22}. These  technologies are promising to meet future demands, but their impact on network planning must be carefully evaluated. Multi-band systems have been studied extensively in a network planning context~\cite{Sambo:multiband, patri:22, MultiBandPower} showing their potential to increase capacity without the need for additional fibers in the network. The impact of probabilistically shaped quadrature amplitude modulation (PS QAM)-capable next-generation BVTs has also been investigated. A trade-off between spectral blockage and minimizing the number of transponders in mesh networks when considering high-baud rate transponders~\cite{Pedro:beyond100G} has been identified and the potential for increased network capacity when exploiting rate-adaptive BVTs has been demonstrated~\cite{ONDM}. Multi-wavelength sources (MWS) are using optical frequency combs to provide several lines with a single laser. Although MWSs introduce a trade-off by lowering the transmitted optical signal-to-noise ratio (\TXOSNR), potential cost-savings due to a reduced number of lasers have been shown \cite{JOCNCombs}.

While operators of data centers for hyper-scale computing may fill large parts of the spectrum at the time of commissioning the network, OTN operators usually follow a pay-as-you-grow strategy, i.e., to maximize their annual revenues, only the equipment needed to provision the requested traffic in the first planning year is acquired, including the optical line system. The purchase of additional transponders to account for traffic growth is delayed until required. 
To enable this strategy, per-period (incremental) planning has been investigated in a multi-layer context~\cite{IncrementalPlanning}, proposing an integer linear programming (ILP) formulation. The joint planning for multiple periods has also been studied \cite{MultiPeriodMILP}, proposing a mixed ILP formulation for cost-efficient multi-period planning. As ILP solutions are too computationally complex for large networks, multi-period planning has also been studied using heuristics, proposing an incremental approach while provisioning the maximum data rate configuration for each lightpath (LP) \cite{patri:22}. 

We propose a proactive multi-period approach to enable OTN operators' pay-as-you-grow deployment strategy. The approach retains the flexibility to react to uncertainty in traffic growth while adding a forward-looking aspect by planning for the final period's estimated traffic in an initial planning step. This approach facilitates the utilization of next-generation optical communications hardware. We show substantial savings of over 60\% in the number of deployed LPs and 12\% of lasers in the considered planning horizon on a national network topology compared to common incremental approaches.

\section{Multi-period Optical Network Planning}
We use heuristics to solve the problem of routing, configuration, and spectrum assignment (RCSA)~\cite{patri:22}. A traffic model, based on the number of internet exchange points and the population at each node \cite{patri:22} is considered. A traffic demand is defined as the aggregate requested traffic (ART) between a source and destination node in the network. For a set of traffic demands, we first sort them in descending order by the shortest path length between the source and destination node. Then, for each demand, we consider the $k$=3-shortest paths. LPs are placed iteratively until the requested data rate is met. Two configuration selection methods are considered: \textit{Just enough}: In case multiple LPs are required to meet the demand, the highest-data rate configuration is selected from a set of configurations that are feasible according to SNR requirements. In case a single LP is sufficient to fulfill the (remaining) requested data rate, the configuration with the lowest bandwidth, that fulfills the data rate requirement, is selected. \textit{Highest}: The highest-data rate configuration of all feasible configurations is selected. Both schemes are followed by the assignment of a spectrum slot to the LPs using first-fit assignment, i.e., the first free slot with sufficient bandwidth is chosen. In case no free spectrum slot is found the LP placement procedure is continued on the next-shortest path. If the provisioned data rate on the 3-shortest-paths is below the requested one, the demand is considered underprovisioned \cite{patri:22}. In case at least three LPs are required to fulfill a demand, placement of a 4-line fixed free spectral range (FSR), i.e., spacing between the lines, MWS is considered. The MWS enables the use of a single laser for transmitting four LPs, coming at the expense of \TXOSNR~penalty compared to single-wavelength transponders~\cite{JOCNCombs}.

To guarantee service level agreements (SLAs) LPs are provisioned according to the estimated end-of-life performance, accounting for aging of equipment and increased non-linear noise in the filled spectrum. The Gaussian-noise model considering inter-channel stimulated Raman scattering (ISRS-GN)~\cite{Buglia:ISRS} is used to estimate the LPs quality of transmission.

\begin{figure}[t]
   \centering
        \includegraphics[width=\linewidth]{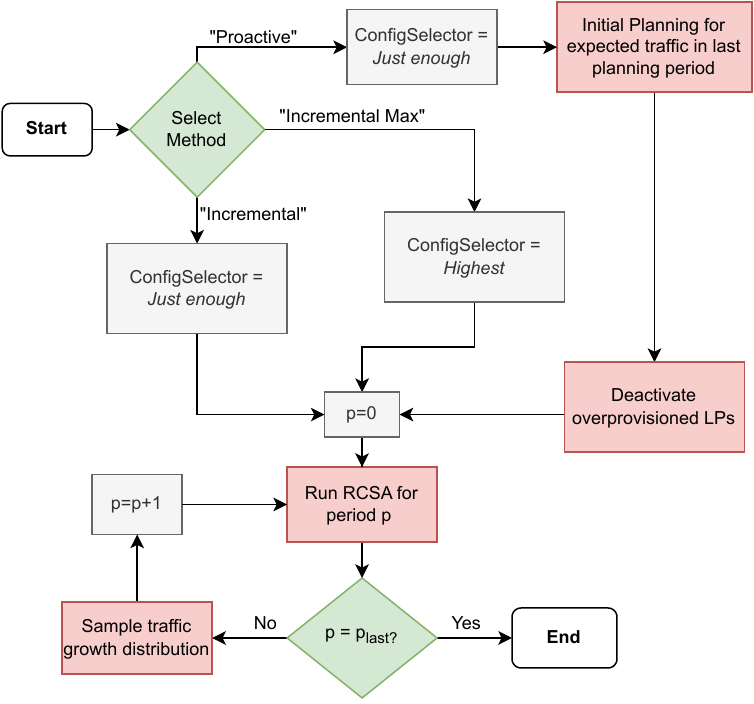}
    \caption{Multi-period planning flow considering the approaches "Proactive", "Incremental" and "Incremental Max".}
    \label{fig:flow}
\end{figure}

We propose a "Proactive" multi-period planning approach as shown in Fig.~\ref{fig:flow}. The presented RCSA using the \textit{Just enough} configuration selection plans the network for the expected traffic in the last considered planning period. To account for uncertainty in traffic growth an overhead (OH) should be considered on top of the traffic estimate. After this initial planning step, we now consider the traffic demands of the first planning period and iterate through all demands and their associated LPs. As soon as LPs data rate meets the demand's requested data rate all additional with the demand associated LPs will be considered inactive. We call this function "Deactivate overprovisioned LPs". Now, for each planning period, the previously described RCSA is employed in the per-period planning stage. The RCSA considers the activation of inactive LPs before LP placement. This approach retains the capability to react to traffic growth uncertainty as additional LPs can be provisioned in the per-period planning stage, in case the actual demand surpasses the estimation considered in the initial planning stage. In each period, only active LPs are required to be provisioned, thereby enabling the operators' preferred pay-as-you-grow deployment strategy. The "Proactive" approach will be compared to the per-period planning approaches "Incremental", using the RCSA with \textit{Just enough} configuration selection, and "Incremental Max" considering \textit{Highest} configuration selection (Fig.~\ref{fig:flow}).

\begin{figure*}[t]
   \centering
    \includegraphics[width=\linewidth]{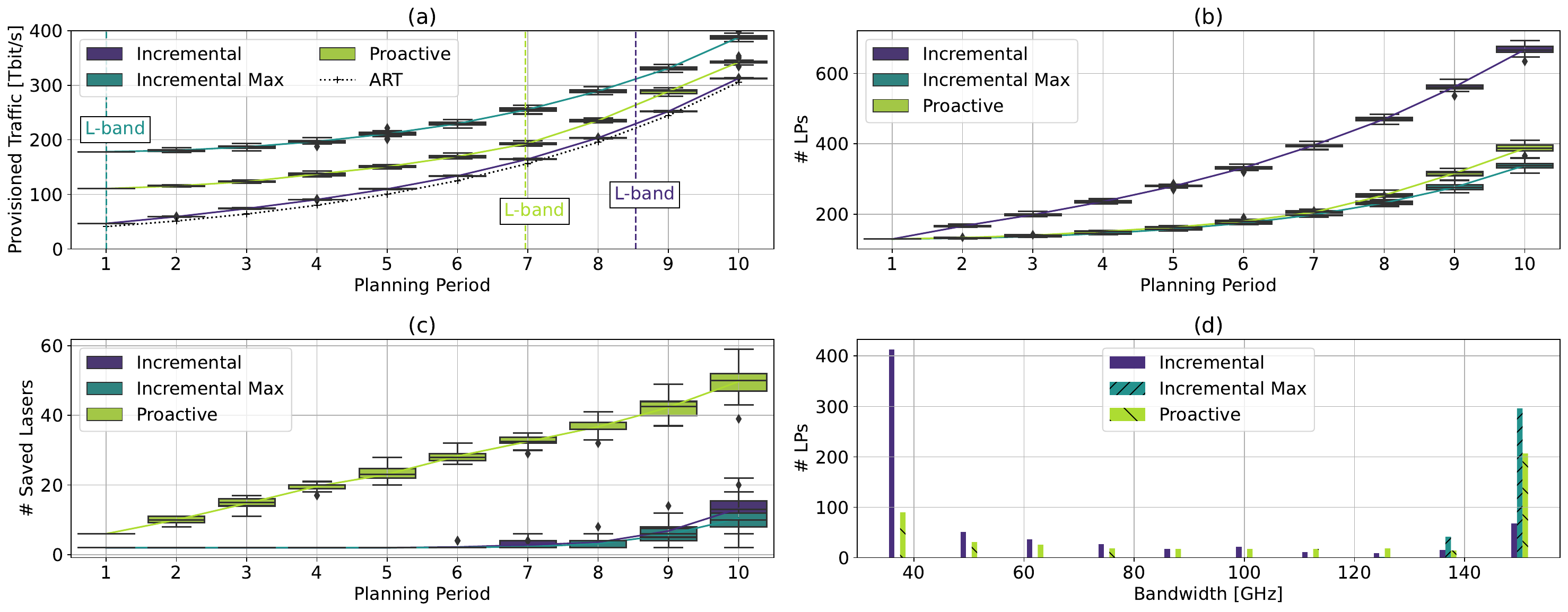}
    \caption{Planning results on Nobel-Germany. (a) Provisioned traffic, (b) provisioned LPs, (c) saved lasers by using MWSs (transmitter side only), (d) bandwidth distribution of deployed configurations in the final planning period.}
    \label{fig:results}
\end{figure*}

\vspace{-0.7\baselineskip}
\section{Network Planning Study}
We assume links in the network consist of 80~km standard single-mode fiber spans followed by an erbium-doped fiber amplifier in the C- and L-band compensating for the attenuation. Per band attenuation and amplifier noise figure values are considered~\cite{patri:22}, as well as frequency-dependent optimized launch powers~\cite{MultiBandPower}. We assume a used bandwidth of 5~THz in each band, with 400 spectrum slots of 12.5~GHz and a guard band of 500~GHz in between. 

The three multi-period planning approaches of "Proactive", "Incremental" and "Incremental Max" are evaluated in terms of provisioned traffic, number of required LPs, and number of saved lasers through the use of MWSs. The considered next-generation rate-adaptive BVTs are assumed to be PS QAM-capable. BVT configurations are assumed to range between 37.5~GHz and 150~GHz slot bandwidth in steps of 12.5~GHz. Therefore, 10 symbol rate options are considered and the relevant modulation rates are chosen according to a configuration pre-selection algorithm~\cite{ONDM}. The considered 4-line Fixed-FSR MWSs are assumed to have a 1~dB \TXOSNR~penalty \cite{JOCNCombs}. The traffic growth of demands is modeled as independent Gaussian random variables with a mean increase of 25\% and a standard deviation of 10\% \cite{patri:22}. For the "Proactive" approach an overhead of 25\% is considered for the initial planning step. The results of 30 realizations of the random traffic growth are presented on the Nobel-Germany topology \cite{sndlib}.

Fig.~\ref{fig:results} (a) shows the provisioned traffic of the different approaches, scaled to the mean aggregate requested traffic (ART). "Incremental Max" is placing almost exclusively LPs with 150~GHz bandwidth. This leads to large overprovisioning and the L-band being deployed already in the first period. For the "Incremental" approach, the provisioned traffic closely follows the ART with minimal overprovisioning while the provisioned traffic for "Proactive" stays in between the two "Incremental" approaches. For 10\% of the samples of the randomized traffic growth, the "Incremental Max" approach leads to minor underprovisioning~\cite{patri:22} of around 1\%. Both "Proactive" and "Incremental" approaches provision all demands for all samples. On average, for "Proactive" the L-band provisioning is postponed to the 7th period with "Proactive" and almost the 9th period with "Incremental". This is due to the trade-off between minimizing the number of required transponders using high-baud rate configurations and spectral blockage also reported in previous work~\cite{Pedro:beyond100G, JOCNCombs}. Increased deployment of high-baud rate configurations and MWSs with "Proactive" leads to a higher spectral utilization than for "Incremental" in the earlier periods. 
Over 60\% more LPs are provisioned with "Incremental" in the last period, as shown in Fig.~\ref{fig:results} (b). Although the lowest number of LPs is required for "Incremental Max", with "Proactive" a significant number of lasers, up to an average of 50 in the last period, are saved through the deployment of MWSs while both "Incremental" approaches save less than 20 lasers on average (Fig.~\ref{fig:results} (c)). For the last planning period, in the "Incremental" approach most LPs are deployed with the minimum bandwidth of 37.5~GHz while "Incremental Max" almost exclusively deploys 150~GHz LPs (Fig.~\ref{fig:results} (d)). For "Proactive", most LPs are deployed with 150~GHz, yet all bandwidth options are utilized, thereby achieving a trade-off between spectral efficiency and minimizing the number of LPs and lasers.

\section{Conclusions}
We present a proactive multi-period planning approach, that enables OTN operators' strategy of pay-as-you-grow network deployment. The approach supports the utilization of next-generation optical communications technology with up to 60\% fewer required LPs and significant savings of 12\% in the number of lasers compared to conventional approaches. 

\section{Acknowledgements}
\footnotesize The work has been partially funded by the German Federal Ministry of Education and Research in the project STARFALL (16KIS1418K).


\printbibliography

@misc{sndlib, title={{SNDlib Problem Instances}}, url={\url{http://sndlib.zib.de/}}, author={{Zuse Institute Berlin}},
 note = {\url{http://sndlib.zib.de/}, Accessed: 2023-01-17}}

@ARTICLE{Emmerich:22,

  author={Emmerich, Robert and Sena, Matheus and Elschner, Robert and Schmidt-Langhorst, Carsten and Sackey, Isaac and Schubert, Colja and Freund, Ronald},

  journal={Journal of Lightwave Technology}, 

  title={{Enabling S-C-L-Band Systems With Standard C-Band Modulator and Coherent Receiver Using Coherent System Identification and Nonlinear Predistortion}}, 

  year={2022},

  volume={40},

  number={5},

  pages={1360-1368},

  doi={10.1109/JLT.2021.3123430}}

@MISC{Jannu, 
title={{Acacia Unveils Industry's First Single Carrier 1.2T Multi-Haul Pluggable Module}},
year= "2022",
author={{Acacia}},
howpublished="https://acacia-inc.com/blog/acacia-unveils-industrys-first-single-carrier-1-2t-multi-haul-pluggable-module/"
}

@ARTICLE{Buglia:ISRS,
  author={Buglia, H. and Jarmolovičius, M. and Vasylchenkova, A. and Sillekens, E. and Galdino, L. and Killey, R. I. and Bayvel, P.},
  journal={Journal of Lightwave Technology}, 
  title={A Closed-Form Expression for the Gaussian Noise Model in the Presence of Inter-Channel Stimulated Raman Scattering Extended for Arbitrary Loss and Fibre Length}, 
  year={2023},
  volume={},
  number={},
  pages={1-10},
  doi={10.1109/JLT.2023.3256185}}

@article{patri:22,
title = {Multi-band transparent optical network planning strategies for 6G-ready European networks},
journal = {Optical Fiber Technology},
volume = {74},
pages = {103118},
year = {2022},
issn = {1068-5200},
doi = {https://doi.org/10.1016/j.yofte.2022.103118},
url = {https://www.sciencedirect.com/science/article/pii/S1068520022003017},
author = {Sai Kireet Patri and Achim Autenrieth and Jörg-Peter Elbers and Carmen Mas-Machuca}}

@article{JOCNCombs,
author = {Jasper Müller and Ognjen Jovanovic and Tobias Fehenberger and Di Rosa, G. and Jörg-Peter Elbers and Carmen Mas-Machuca},
journal = {J. Opt. Commun. Netw.},
title = {{Multi-Wavelength Transponders for High-capacity Optical Networks: A Physical-layer-aware Network Planning Study}},
year = {2023},
doi = {10.1364/JOCN.483320}}

@ARTICLE{Pedro:beyond100G,
  author={Pedro, Joao and Costa, Nelson and Pato, Silvia},
  journal={Journal of Optical Communications and Networking}, 
  title={{Optical Transport Network Design Beyond 100 GBaud [Invited]}}, 
  year={2020},
  volume={12},
  number={2},
  pages={A123-A134},
  doi={10.1364/JOCN.12.00A123}}

@INPROCEEDINGS{ONDM,
  author={Jasper Müller and Di Rosa, G. and Tobias Fehenberger and Mario Wenning and Sai Kireet Patri and Jörg-Peter Elbers and Carmen Mas-Machuca},
  booktitle={27th International Conference on Optical Network Design and Modelling (ONDM)}, 
  title={{On the Benefits of Rate-Adaptive Transceivers: A Network Planning Study}}, 
  year={2023},
  doi={}}

@INPROCEEDINGS{COMBS,
  author={Di Rosa, G. and Frank Smyth and Deseada Gutierrez, M. and Jasper Müller and Benjamin Wohlfeil, and Jörg-Peter Elbers},
  booktitle={Signal Processing in Photonic Communications (SPPCom)}, 
  title={{Advancements and Applications of Comb-Based Transceivers in Coherent Optical Networks}}, 
  year={2023},
  doi={}}

@ARTICLE{MultiBandPower,
  author={Correia, Bruno and Sadeghi, Rasoul and Virgillito, Emanuele and Napoli, Antonio and Costa, Nelson and Pedro, Joao and Curri, Vittorio},
  journal={Journal of Optical Communications and Networking}, 
  title={{Power control strategies and network performance assessment for C+L+S multiband optical transport}}, 
  year={2021},
  volume={13},
  number={7},
  pages={147-157},
  doi={10.1364/JOCN.419293}}

@ARTICLE{Jiang:19,
  author={Jiang, Wei and Han, Bin and Habibi, Mohammad Asif and Schotten, Hans Dieter},
  journal={IEEE Open Journal of the Communications Society}, 
  title={The Road Towards 6G: A Comprehensive Survey}, 
  year={2021},
  volume={2},
  number={},
  pages={334-366},
  doi={10.1109/OJCOMS.2021.3057679}}

@ARTICLE{MultiPeriodMILP,
  author={Mesquita, Leonardo A. J. and Assis, Karcius D. R. and Almeida, Raul C.},
  journal={The Journal of Supercomputing}, 
  title={Multi-period traffic on elastic optical networks planning: alleviating the capacity crunc}, 
  year={2021},
  volume={77},
  number={6},
  pages={5468-5491},
  doi={10.1007/s11227-020-03493-7}}

@inproceedings{Sambo:multiband,
author = {Sambo, Nonhlanhla and Ferrari, Alessio and Napoli, Antonio and Costa, Nelson and Pedro, Joao and Sommerkohrn-Krombholz, B. and Castoldi, Piero and Curri, Vittorio},
year = {2019},
month = {01},
pages = {53 (4 pp.)-53 (4 pp.)},
title = {{Provisioning in multi-band optical networks: a C+L+S-band use case}},
doi = {10.1049/cp.2019.0787}
}

@INPROCEEDINGS{IncrementalPlanning,
  author={Papanikolaou, P. and Christodoulopoulos, K. and Varvarigos, E.},
  booktitle={2017 International Conference on Optical Network Design and Modeling (ONDM)}, 
  title={Incremental planning of multi-layer elastic optical networks}, 
  year={2017},
  volume={},
  number={},
  pages={1-6},
  doi={10.23919/ONDM.2017.7958534}}

\vspace{-4mm}

\end{document}